\documentclass[newclass, twocolumn,proceeding]{rmaa}
      
\title{Compact Binaries in Globular Clusters}
\author{Natalia Ivanova and Frederic A.\ Rasio
  \affil{Northwestern University, USA}
}    

\fulladdresses{
 \item Natalia Ivanova: Dept of Physics and Astronomy, Northwestern
University, 2145 Sheridan Rd, Evanston, IL 60208, USA ({\tt
nata@northwestern.edu}).
 \item Frederic A.\ Rasio: Dept of Physics and Astronomy, Northwestern
University, 2145 Sheridan Rd, Evanston, IL 60208, USA
({\tt rasio@northwestern.edu}).
}

\shortauthor{Ivanova and Rasio} \shorttitle{Globular Clusters}

\ReceivedDate{} 
\AcceptedDate{} 
\SetYear{2004}

\abstract{In dense stellar systems the frequent dynamical interactions
between
stars play a crucial role in the formation and evolution of compact
binaries. We study these processes using a novel approach combining a
state-of-the-art binary population synthesis code with a simple treatment of
dynamical interactions in dense star cluster cores.
Here we focus on the dynamical and evolutionary processes
leading to the formation of compact binaries containing white dwarfs in dense globular clusters.
We demonstrate that dynamics can increase by factors $\sim 2 - 100$
the production rates of interesting binaries
such as cataclysmic variables, ``nonflickerers''
(He white dwarfs with a heavier
dark companion), merging white dwarf binaries with total masses above the
Chandrasekhar limit, and white dwarf binaries emitting gravitational waves in the LISA band.}


\begin{document}
\maketitle

\section{Introduction}

From the earliest observations of X-ray binaries in globular clusters
it has always been clear that they must be
very efficient factories for the production of compact
binary systems \citep{1975ApJ...199L.143C}.
The overabundance of compact binaries in clusters, as compared to the field,
must be a result of close stellar encounters.
The key processes that affect the binary population in dense cluster environments 
include the destruction of wide binaries, hardening of close binaries 
(following ``Heggie's law'' \citep{Heggie_75}),
and exchange interactions, through which low-mass companions tend to be replaced
by a more massive participant in the encounter. 
As a result of these processes, in the dense cores of globular clusters, binaries 
are strongly depleted and 
their period distribution is very different from that of a field population 
(Ivanova et al.\ 2004).
These processes also lead to an interesting and complex interplay between dynamics and binary
evolution. For example, exchange interactions involving compact objects often produce systems
that will evolve through a common-envelope (CE) phase and form very short-period
binaries, which are much less common in field populations \citep{2000ApJ...532L..47R}.

There are two possible approaches to the study of binary evolution and
dynamics in globular clusters.
One can start from $N$-body simulations and introduce various simplified
treatments of binary star evolution. This has been the traditional 
approach for many years \citep[for recent examples 
see][]{SharHurley_2002,2001MNRAS.321..199P}.
Alternatively, one can start from a binary population synthesis code
and add a treatment of dynamical interactions. This approach
was pioneered by Portegies Zwart et al.\ (1997) and has been adopted in our
recent work.
It has the great advantage that it is computationally much less expensive
than $N$-body simulations, so
that more exploration of the (enormous) parameter space is possible, and more
realistic simulations, using sufficiently large numbers of stars and
binaries, are possible today. In contrast, even when using special-purpose
GRAPE computers, $N$-body simulations are still limited to
 smaller systems like open clusters with limited coverage of parameter space
and with unrealistically
small numbers of binaries \citep[see][]{BinFrac_04, 2003MNRAS.343.1025W}.

In our code we combined {\tt StarTrack}, a state-of-the-art  binary
population synthesis code
\citep{Chris_02} and {\tt FewBody}, a small-$N$-body integrator that we use  
to compute 3-body and 4-body interactions
\citep{Fregeau_FB2_04,Fregeau_FB1_04}.
Currently we adopt a simple two-zone model, in which the cluster is partitioned
into an inner core and an outer halo, 
with all interactions assumed to take place in the core.
This background cluster model remains 
unchanged throughout the evolution \citep{hut_1992}. In particular, the core
density is assumed constant.
However, our ultimate aim is to incorporate full dynamical Monte Carlo models
\citep{2003ApJ...593..772F}.
In a typical simulation we start with  $N\sim 10^5$ stars, with
between 50\% and 100\% binaries. This high primordial binary fraction 
(much higher than assumed in all previous studies) is needed in order to
match the observed binary fractions in globular cluster cores today
\citep{BinFrac_04}.

\section{Compact Binaries and Mergers}

\begin{figure}
\begin{center}
\includegraphics[width=\columnwidth]{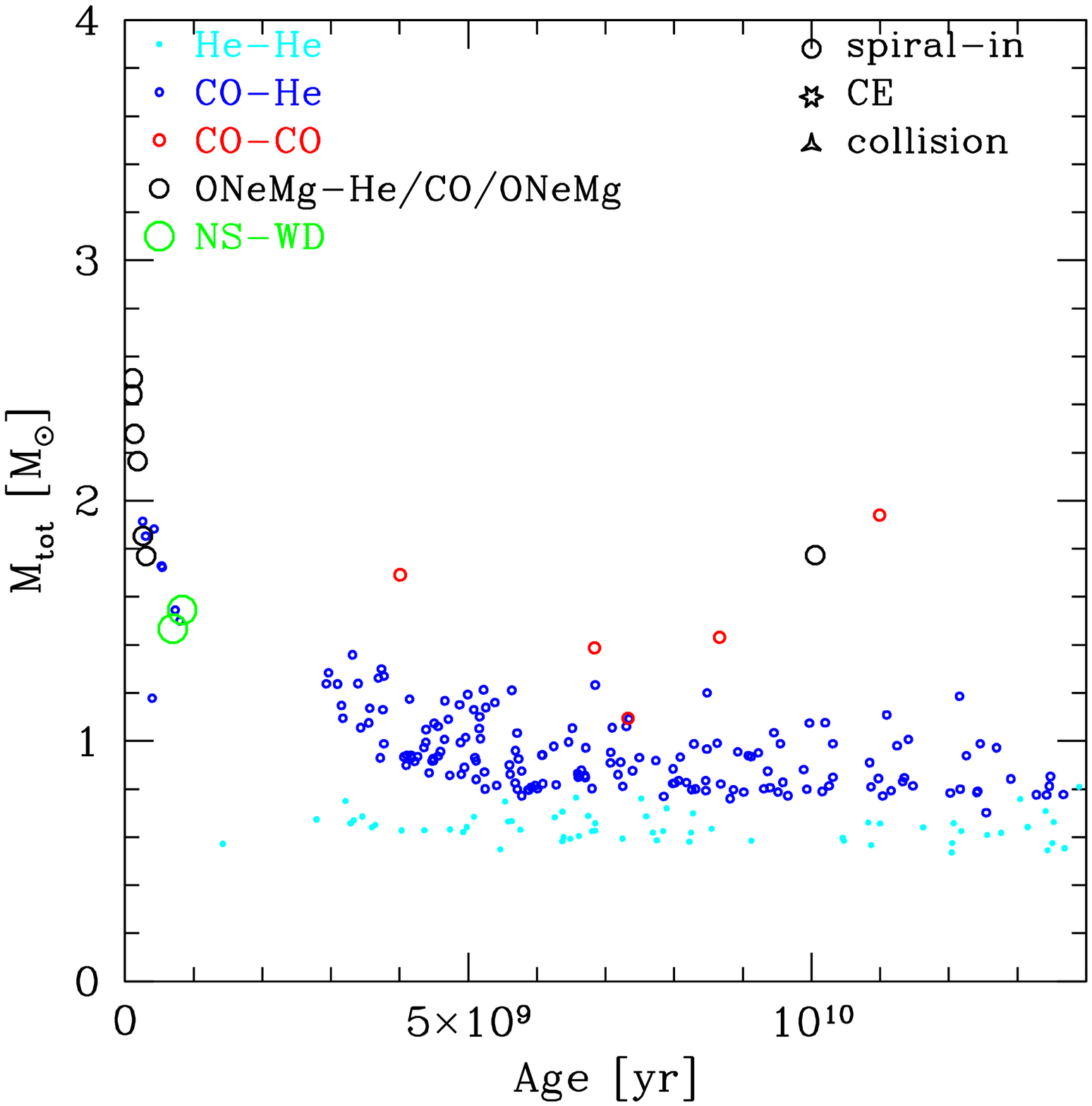}
\includegraphics[width=\columnwidth]{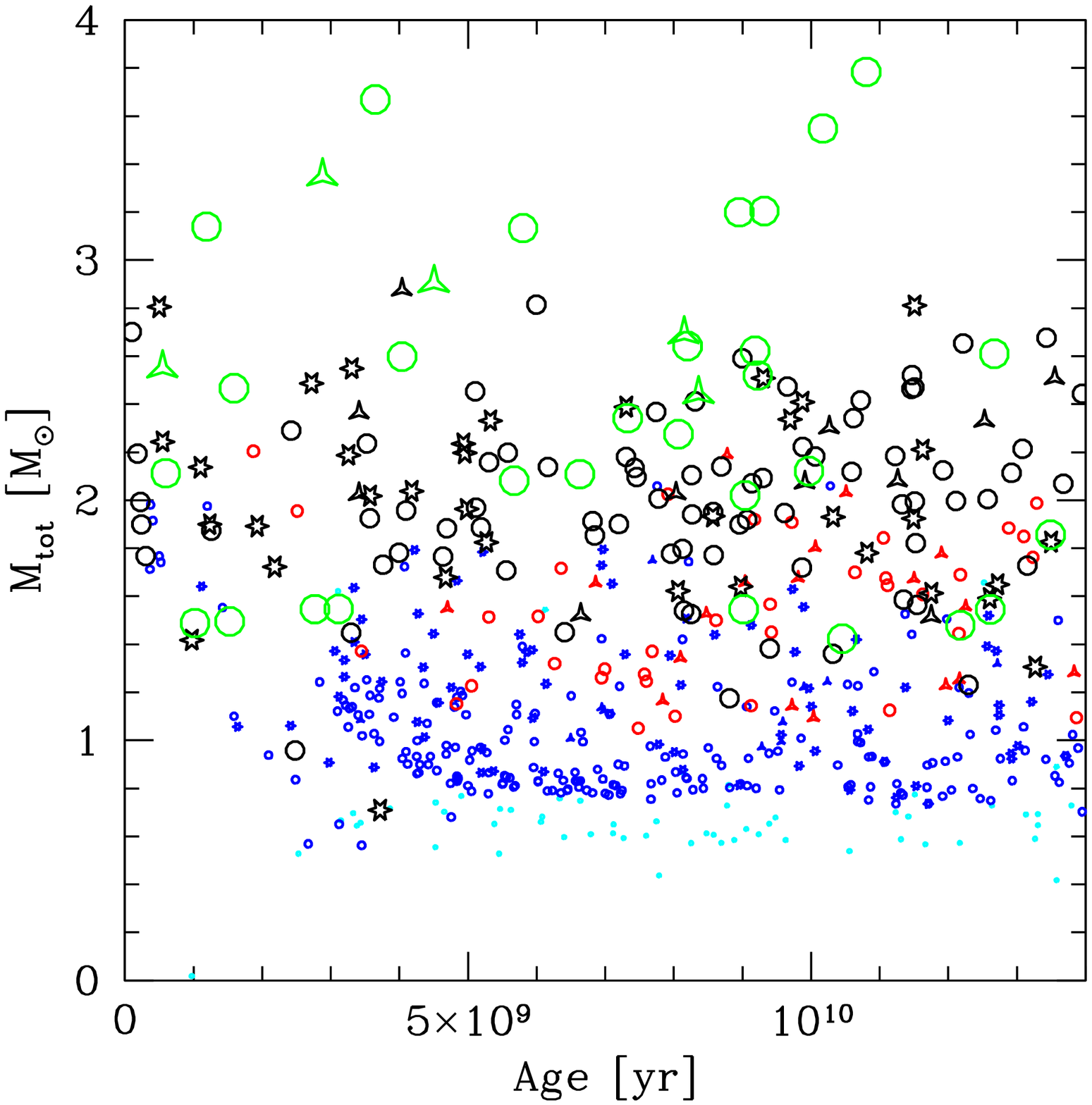}
\caption{
	Mergers and dynamical formation of compact binaries containing white dwarfs (WD) or
	neutron stars (NS). The horizontal axis gives the cluster age when the event
	occurred and the vertical axis gives the total mass of the binary.
	The top panel shows a population evolved without dynamical interactions, 
	while the bottom panel shows the same initial population evolved in a
	dense cluster with core density $n_c=10^5\,$pc$^{-3}$.
	 The size of each symbol indicates the types of stars involved, 
     while the shape indicates the type of event: binary merger 
     (round dots; usually driven by
	gravitational radiation or spiral-in during a common envelope phase), 
	compact binary formation following a 
	common envelope phase (stars), 
	or merger following a physical collision (triangles). 
	It is clear that dynamics increases both the total number of events and the
	typical mass of mergers and compact binaries.
	}
\end{center}
\end{figure}

Consider the evolution of a typical dense cluster with central
velocity dispersion $\sigma=10\,{\rm km}\,{\rm s}^{-1}$ and core density
$n_{\rm c}= 10^5\,$pc$^{-3}$. Dynamical interactions lead to greatly
enhanced numbers of compact binaries containing white dwarf (WD) and neutron star (NS)
components, and, in particular, to much larger numbers of heavier compact
binaries (Fig.~1).
Here we focus in particular on the fate of WDs involved in CE events 
leading to compact binary formation or 
mergers of WDs driven by gravitational wave emission.
Our simulations confirm that the formation rate of compact binaries
containing a Helium WD with a heavier companion via CE events is increased significantly in dense clusters. The brighter Helium WDs in these binaries 
could be detectable as ``non-flickerers.'' These were observed for the 
first time in the core of NGC 6397 \citep{Cool_1998} and are indeed thought
to be double WD binaries containing a young Helium WD with an older and heavier WD 
companion \citep{HKR_2003}.

We can also examine the rate of double WD mergers, and, in particular, 
those for which the total mass is $\ge M_{\rm Ch}\simeq 1.4 M_\odot$).
These so-called supra-Chandrasekhar mergers could lead 
either to a Type~Ia supernova, or to a ``merger-induced''
collapse of the remnant to form a NS (and perhaps a millisecond radio pulsar). 
It is possible that the NS 
in this case is formed without a significant kick 
(see the article by Podsiadlowski in this volume).
An increased rate of Type~Ia supernovae from star cluster dynamics would likely
be redshift-dependent (as more stars are formed in starburst environments ---
which favor star cluster formation --- at higher redshifts) and this has
important potential consequences for their use in cosmology  
\citep[for a review see][]{2001ARA&A..39...67L}. Alternatively,
merger-induced collapse could lead to the formation of neutron stars and
millisecond pulsars in clusters, thereby alleviating or perhaps eliminating the
NS ``retention problem'' (the ejection of most NS from the shallow cluster
potentials if they are born with the natal kicks expected from asymmetric
supernova explosions; see, e.g., Chen \& Leonard 1993; Pfahl, Rappaport,
\& Podsiadlowski 2002)

\begin{figure}
\begin{center}
\includegraphics[width=\columnwidth]{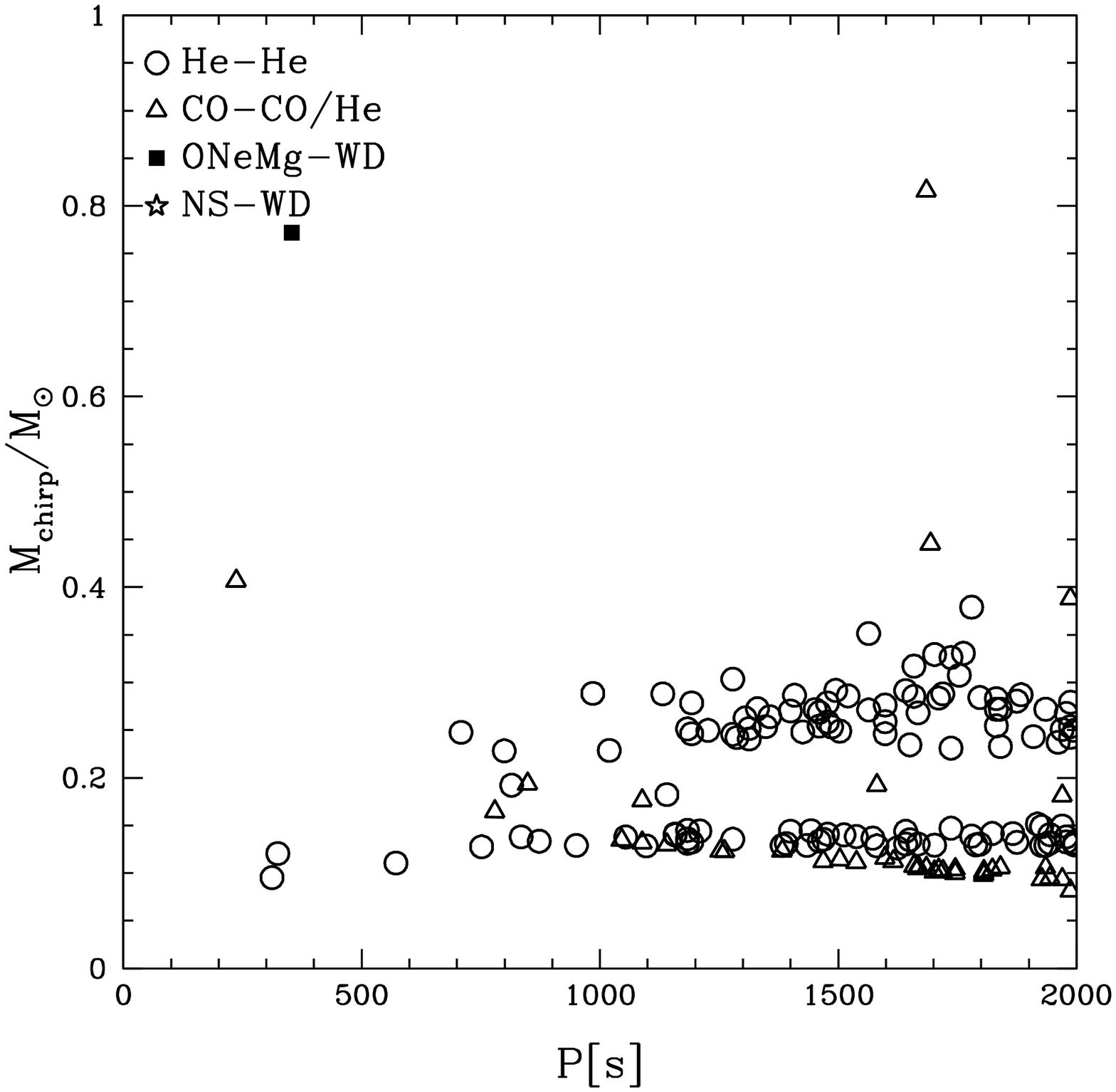}
\includegraphics[width=\columnwidth]{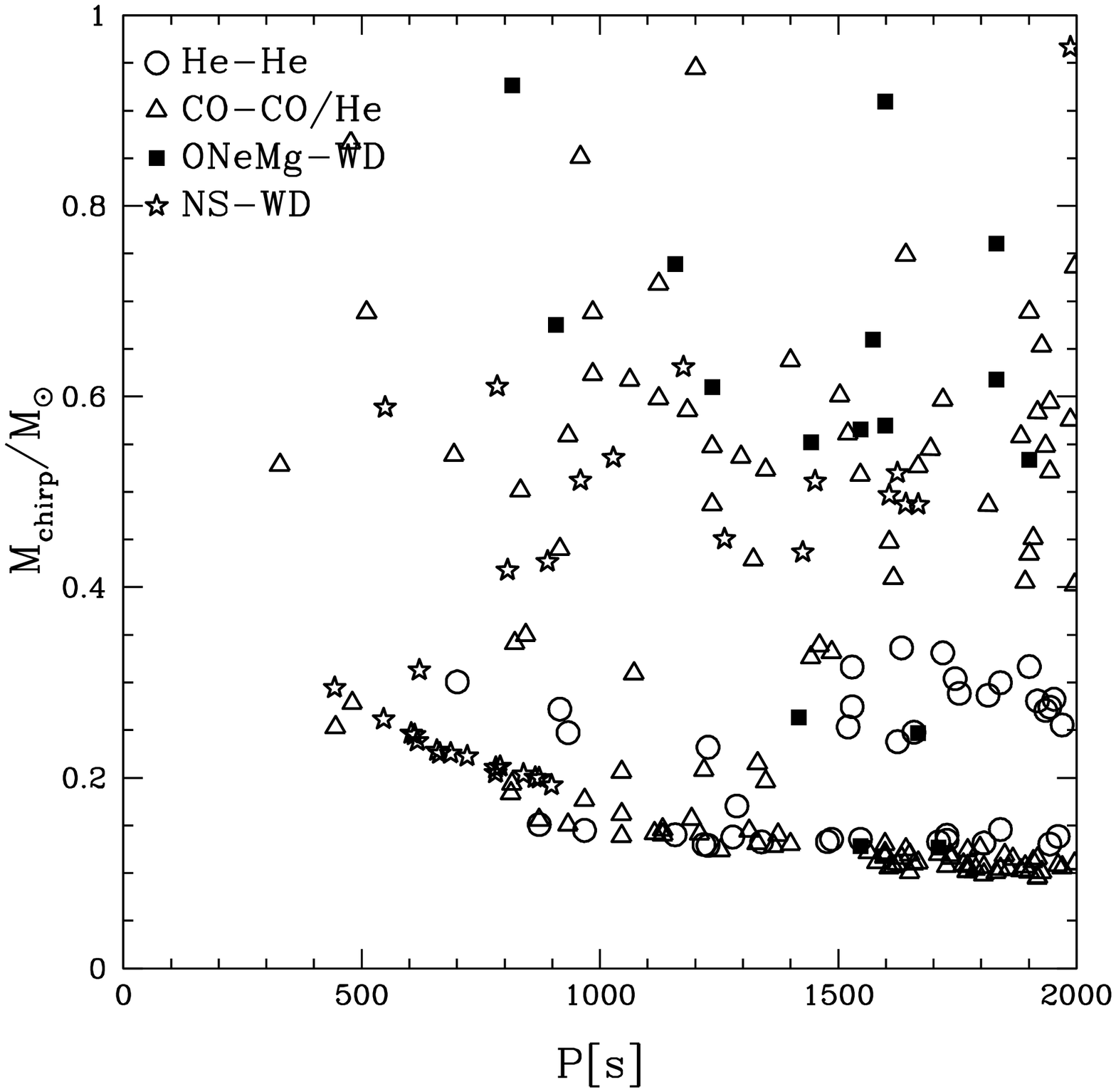}
\caption{
	All LISA-detectable binaries during the last Gyr of evolution for a
	typical dense cluster model  
	with core density $n_c=10^6\,$pc$^{-3}$. 
	Each source is shown only once, at a randomly chosen moment 
	in its evolution across the LISA band. Here
	$P$ is the binary orbital period and
	$M_{\rm chirp}= (M_1M_2)^{3/5}/(M_1+M_2)^{1/5}$ is the chirp mass.
	The different symbols indicate binaries with different types of components.
	We again compare a field population (top) to our cluster model 
	including dynamical interactions  (bottom). 
	}
\end{center}
\end{figure}

The enhanced production rate of double WD mergers in dense stellar clusters
was first discussed in detail by \citet{SharHurley_2002}.
They estimated that, for stars born in open clusters (which can be simulated
directly using their $N$-body code), 
the supra-Chandrasekhar WD merger rate can be increased by an order of magnitude.
The results of \citet{SharHurley_2002} are based on 
$N$-body simulations for a typical open cluster 
containing $\sim10^4$ stars with 10\% primordial binary fraction,
and with
$\sigma=2\,{\rm km}\,{\rm s}^{-1}$ and $n_{\rm c}= 10^3\,$pc$^{-3}$.
Our simulations are for much denser and massive star clusters with a
primordial binary fraction of 100\%. Not surprisingly, we find an 
even larger number of supra-Chandrasekhar WD mergers, with an enhancement 
factor (compared to a field population) closer to $\sim 100$ for a typical dense globular cluster. 
In contrast, the {\em total\/} number of double WD mergers (of all types) is typically
increased by a factor of a few only. The majority of these mergers are driven
ultimately by gravitational radiation, although
a few come from physical collisions of WDs during hard binary 
encounters (Fregeau et al.\ 2004).

\section{LISA Sources}

Our results show that compact double WD binaries are mainly formed dynamically 
and have typically experienced
multiple hardening encounters before merging.
Prior to merger, these systems may be detectable as gravitational wave sources
by LISA, when their orbital
period becomes smaller than about $2000\,$s 
(Benacquista, Portegies Zwart \& Rasio, 2001).
This limit on the orbital period is imposed by the background noise
from Galactic binaries. At the same time, the positional accuracy of 
LISA is much greater for binaries with these shorter periods, so that the sources
can then be associated with specific globular clusters in our Galaxy. 

Figure~2 shows all LISA-detectable sources (chirp masses and periods)
that appeared during the last Gyr of our simulation for a typical dense cluster.
For this model, where the total cluster mass today is about
$2\times10^5\,M_\odot$, at least one LISA source is present at any given time.
On average, there are $\sim 5$ LISA sources at any given moment,
about twice the number predicted for a field population.
In addition to this rise in the number of sources, we also note changes
in their typical properties: in particular,
the number of sources with chirp mass above $0.4\,M_\odot$
is increased significantly in the cluster model.
We also find that NS-WD binaries represent about 20\% of all
LISA sources.
Therefore, the number of LISA-detectable NS-WD binaries 
in all Galactic globular clusters ($M_{\rm tot}\sim 10^{7.5}M_\odot$)
could be as high as $\ga 100$, while the total number of detectable
WD-WD binaries could be $\ga 500$.
However, since the cluster models we have used so far in our simulations 
are
denser than average, these numbers should be taken as upper bounds.

\section{Mass Transfer Systems}

As they spiral in and evolve across the LISA band, WD-WD and NS-WD binaries
will eventually come into contact around a period $P\sim 100\,$s.
At first the binary orbit shrinks as it looses angular momentum to
gravitational radiation. However, during stable mass transfer,
the  orbit will evolve towards a larger period. 
During this mass transfer phase the \adjustfinalcols binary can also appear as 
an ultracompact X-ray binary (NS-WD) or an AM CVn type cataclysmic variable (WD-WD).
As expected, we find that the population of these mass transfer binaries
is also increased significantly by dynamical interactions. For example,
for our typical cluster model with $n_c=10^5\,$pc$^{-3}$, we predict 
 about 50 AM CVn binaries, 
roughly twice the number obtained without dynamics.

\medskip

This work was supported by NASA ATP Grant NAG5-12044 and a Chandra Theory
grant. F.A.R.\ thanks the organizers of IAU Colloquium 194 for support and
acknowledges the hospitality of
the Kavli Institute for Theoretical Physics.


\begin{thebibliography}

\bibitem[Belczynski et~al.\ (2002)]{Chris_02}
{Belczynski}, K., {Kalogera}, V., \& {Bulik}, T. 2002, \apj, 572, 407

\bibitem[Benacquista, Portegies Zwart \& Rasio (2001)]{BZR2001}
Benacquista, M.J., Portegies Zwart, S., and Rasio, F.A. 2001, Class.\
Quantum Grav., 18, 4025

\bibitem[Chen \& Leonard (1993)]{}
Chen, K., \& Leonard, P.J.T. 1993, ApJ, 411, L75

\bibitem[Clark(1975)]{1975ApJ...199L.143C} Clark, G.W.\ 1975, \apjl, 199,
L143 

\bibitem[Cool et al.(1998)]{Cool_1998} Cool, A.M., Grindlay,
J.E., Cohn, H.N., Lugger, P.M., \& Bailyn, C.D. 1998, \apjl, 508, L75

\bibitem[Fregeau et al.\ (2003)]{2003ApJ...593..772F}
Fregeau, J.M., G{\" u}rkan, M.A., Joshi, K.J., \& Rasio, F.A.\ 2003,
\apj, 593, 772

\bibitem[{{Fregeau} \& {Rappaport}(2004)}]{Fregeau_FB1_04}
{Fregeau}, J.M., \& {Rappaport}, S.A. 2004, in preparation

\bibitem[Fregeau et~al.\ (2004)]{Fregeau_FB2_04}{Fregeau}, J.M., {Cheung},
P.,
{Portegies Zwart}, S.F., \& {Rasio}, F.A.
2004, submitted to \mnras\ [astro-ph/0401004]
 
\bibitem[Hansen, Kalogera, \& Rasio(2003)]{HKR_2003} Hansen,
B.M.S., Kalogera, V., \& Rasio, F.A.\ 2003, \apj, 586, 1364

\bibitem[Heggie (1975)]{Heggie_75}
Heggie,D.C. 1975, \mnras, 173, 729
 
\bibitem[Hut, McMillan, \& Romani(1992)]{hut_1992} Hut, P.,
McMillan, S., \& Romani, R.W. 1992, \apj, 389, 527

\bibitem[Ivanova et al.\ (2004)]{BinFrac_04}
Ivanova, N., {Belczynski}, K., Fregeau, J.M.,\& {Rasio}, F.A. 2004, submitted
to \apjl  [astro-ph/0312497]

\bibitem[Leibundgut(2001)]{2001ARA&A..39...67L} Leibundgut, B.\ 2001, 
\araa, 39, 67 

\bibitem[Pfahl, Rappaport \& Podsiadlowski (2002)]{}
Pfahl, E., Rappaport, S., \& Podsiadlowski, P. 2002, ApJ, 573, 283

 \bibitem[Portegies Zwart, Hut, McMillan, \& Verbunt(1997)]
{ecology_ii} Portegies Zwart, S.F., Hut, P.,
McMillan, S.L.W., \& Verbunt, F.\ 1997, \aap, 328, 143

\bibitem[Portegies Zwart, McMillan, Hut, \&
Makino(2001)]{2001MNRAS.321..199P} Portegies Zwart, S.F., McMillan,
S.L.W., Hut, P., \& Makino, J.\ 2001, \mnras, 321, 199

\bibitem[Rasio, Pfahl, \& Rappaport(2000)]{2000ApJ...532L..47R} Rasio,
F.A., Pfahl, E.D., \& Rappaport, S.\ 2000, \apjl, 532, L47

\bibitem[Shara \& Hurley(2002)]{SharHurley_2002}
Shara, M.M.~\& Hurley, J.R.\ 2002, \apj, 571, 830

\bibitem[Wilkinson et al.(2003)]{2003MNRAS.343.1025W} Wilkinson, M.I.,
Hurley, J.R., Mackey, A.D., Gilmore, G.F., \& Tout, C.A.\ 2003, \mnras,
343, 1025 

\end{thebibliography}
\end{document}